\documentclass[aps,prc,preprint,tightenlines,groupedaddress,showpacs]{revtex4}
\usepackage{graphicx}
\begin{document}


\title{\bf Density dependence of isospin observables
in spinodal decomposition}

\author{M. Colonna}
\affiliation{{\small\it Laboratori Nazionali del Sud,
        Via S. Sofia 62, I-95123, Catania, Italy}}
\author{F. Matera}
\email{matera@fi.infn.it}
\affiliation{{\small\it Dipartimento di Fisica, Universit\`a degli
Studi di Firenze,}}
\affiliation{
{\small\it Istituto Nazionale di Fisica Nucleare, Sezione di
Firenze,}\\
{\small\it Via G. Sansone 1, I-50019, Sesto F.no (Firenze), Italy}}


\begin{abstract}

Isotopic fluctuations in fragment formation are investigated in 
a quasi--analytical description of the
spinodal decomposition scenario. By exploiting the
fluctuation--dissipation relations the covariance matrix of density 
fluctuations is derived as a function of the wave vector ${\bf k}$ 
for nuclear matter at given values
of density, charge asymmetry, temperature, and of the time that the
system spends in the instability region. Then density fluctuations
in ordinary space are implemented with a Fourier transform performed
in a finite cubic lattice. Inside this box, domains with different 
density coexist, from which  clusters of nucleons eventually emerge.
Within our approach, the isotopic distributions are determined 
by the $N/Z$ ratio of the leading unstable isoscalarlike mode 
and by isovectorlike fluctuations present in the matter undergoing 
the spinodal decomposition. Hence the average value of the  
$N/Z$ ratio of clusters and the width of the relative distribution 
reflect the properties of the symmetry energy. 
Generating a large number of events, these calculations allow a
careful investigation of the cluster isotopic content as a function 
of the cluster density. A uniform decrease of the average charge 
asymmetry and of the width of the isotopic distributions 
with increasing density is observed. Finally we remark that the results
essentially refer to the early break--up of the system.
\end{abstract}

\pacs{21.65.Cd, 21.65.Ef, 24.60.Ky, 25.70.Pq, 21.60.Jz}
\maketitle

\section{Introduction}

Reactions with charge--asymmetric systems open the possibility to
learn about the properties of the symmetry term of the nuclear
interaction in conditions of density and temperature away from the
ordinary values. In particular, the study of multi-fragmentation
mechanisms in neutron--rich systems should allow to get information on the
behaviour of the symmetry energy in density regions below
saturation, where the nuclear system may undergo a liquid--gas phase
transition. Constraints on the form of the density dependence
of the symmetry energy are important not only for a better knowledge
of the nucleon---nucleon interaction, and hence its extrapolation to
the structure of exotic nuclei \cite{Bro00,Hor01,Fur02}, but also
for the study of the neutron star crust, and of
supernova explosions, where a key issue is the clustering of
low--density matter \cite{Lat91,Pet95,Lee96,Lat07}.

The dynamics of first--order phase transitions is often induced by
instabilities against fluctuations of the order parameter.   In
dissipative heavy ion collisions, nuclear matter may be pushed
inside the coexistence region of the nuclear liquid-gas phase
diagram. Then, the observed abundant fragment formation may take
place through a rapid amplification of spinodal instabilities.
Experimental results pleading in favor of such a spinodal
decomposition have recently been reported \cite{Tab06,Fra01}. Spinodal
instabilities in charge-asymmetric systems have been widely
investigated from the theoretical point of view. The most important
effect induced by the charge asymmetry is the so-called isospin
distillation: fragments (liquid) appear more symmetric with respect
to the initial matter, while light particles (gas) are more
neutron--rich \cite{Mue95,Bao97,Bar98,Rep04,Rep05,Xu00}. The amplitude of this
effect depends on specific properties of the isovector part of the
nuclear interaction, namely on the value and the derivative of the
symmetry energy at low density \cite{EPJA06}. Moreover, apart from
the distillation effect, that determines the average fragment
isotopic composition, the symmetry energy value influences also the
width of the isotopic distributions. This feature has been recently
exploited in the so--called isoscaling analysis \cite{Tsa01}, 
where information on the symmetry energy behaviour 
is extracted from the study of ratio
of the isotope yields obtained from two reactions with different
charge asymmetry \cite{Bot02,Buy05,Igl06,Sou07}.\par 
In this paper we focus on a detailed study of  
isotopic properties of nucleon clusters, 
as obtained within the spinodal decomposition scenario.
In order to select this mechanism, we consider nuclear matter
initialized at a given temperature and at low density, inside a box
with periodic boundary conditions, under the action of a stochastic
field self--consistently determined \cite{Col05,Mat00}. In this case,
the normal modes of the density fluctuations are plane waves. The
instabilities are treated in the linear approximation, i.e.
retaining only first--order terms in the stochastic field or in the
density fluctuations. This leads to a quasi--analytical description
of the growth of instabilities
and the consequent formation of clusters of nucleons, that can be
considered as excited primary fragments. 
For each considered system, a large variety of possible outcomes may
be obtained, according to the initial density fluctuation values. In
this way it is possible to collect a large number of events with
much reduced computational effort, allowing to perform  a  thorough
analysis of the isotopic content of nucleon clusters, the isospin
distillation and isotopic distributions, in connection with the
ingredients of the effective nuclear interaction employed.

We would like to stress that our analysis refers essentially to
the early break-up of the system, where excited nucleon clusters,
reflecting the nuclear matter properties, can be recognized. At this level,
shell effects, that influence the N/Z ratio of real nuclei, are not
considered. In particular, we will discuss the relation 
between isotopic properties and density of the primary fragments 
(at the moment of fragment formation). In fact, in the spinodal 
decomposition scenario, domains with different density values, 
from which clusters of nucleons eventually
emerge, coexist. Clusters of intermediate mass, that may originate
from the vapour phase or from the liquid phase as well,  may have
different isotopic properties, depending on the density value of the
domain where they are formed. For instance, the isospin distillation
mechanism, that is a feature of isoscalarlike oscillations, is more
and more effective as soon as the density gets larger in the domain
considered.

In the actual disassembly of a nuclear system (in central nuclear
collisions) the vapour phase is generally formed at the surface of
the system, where also a larger radial collective flow is observed.
Hence one could expect a relation between fragment isotopic
observables and kinematical properties \cite{ColBa07}. In particular, as
we will discuss in the following, clusters emerging from low density
regions should be more neutron--rich and with wider isotopic
distributions. These features are peculiar of the spinodal
decomposition scenario and of the occurrence of first--order
liquid--gas transitions. Hence, in addition to the information that
can be gained about the low--density behaviour of the symmetry
energy, this investigation would allow to shed some light on the
fragmentation mechanism itself. 

The paper is organized as it follows. In Sec. II we outline the
formalism developed in Refs. \cite{Col05,Mat00} for infinite nuclear
matter and its implementation in a finite cubic lattice. In Sec. III
we discuss the results of our calculations. Finally, in Sec. IV a
summary and conclusions are given.

\section{\label{} Formalism}
In this section we outline the main steps of the general formalism
developed in Refs. \cite{Col05,Mat00} to evaluate fluctuations of the
one--body density for asymmetric nuclear matter inside the unstable
(spinodal) region of the nuclear matter phase diagram. In our
approach fluctuations of the proton and neutron densities are
induced by a stochastic field which couples with the costituents of
the system. By means of the fluctuation--dissipation theorem the
stochastic field is self--consistently determined. For nuclear
matter in the spinodal region the time growth of
fluctuations is essentially due to the unstable mean field.
Therefore we neglect the effects of nucleon-nucleon collisions in
the time evolution of fluctuations. Collisions would mainly add
a damping to the growth rate of the fluctuations and should not
change the main results of our calculations, at least at a
qualitative level. \par 
The approach of Refs. \cite{Col05,Mat00} has
reference to infinite nuclear matter, thus normal modes are plane
waves associated with wave numbers ${\bf k}$. The growth of
fluctuations leads to the formation of large density domains, that
can be associated with fragments. In Refs. \cite{Col05,Mat00} the
fragment recognition  is based on a procedure which allows to
determine the probability distribution of the domains containing
correlated density fluctuations. We have then identified the pattern
of correlated domains with the fragmentation pattern. In such a way,
we could make predictions on the distributions of clusters. This
procedure is based on an ``ansatz'' which relates the Gaussian
distributions for the different modes of density fluctuations to the
probability distribution of the correlated domains in the ordinary
space.

In the present paper we follow a different strategy. From the
probability distributions of fluctuations in the ${\bf k}$ space,
induced by the stochastic field,  we directly generate a certain
number of events characterized by stochastic distributions of
density fluctuations in the coordinate space. This is accomplished
by performing a Fourier expansion in a finite cubic lattice of
volume $V$.

In this way a substantial improvement of the approach of Refs. 
\cite{Col05,Mat00} is obtained. Event by event analyses can be
performed. Moreover, one can investigate new features of the
fragmentation process, such as, for instance, the isotopic content of 
nucleon clusters as a function of their density.

\subsection{Density fluctuations}
The key quantity for the evaluation of the density fluctuations is
the density--density response function. In a linear approximation
for the stochastic field, the Fourier transform of the response
function is given by the equation \cite{Col05,Mat00}:
\begin{equation}
D_{i,j}(k,\omega)=\,D^{(0)}_i(k,\omega)\delta_{i,j}+\Sigma_lD^{(0)}_i(k,\omega)
{\cal A}_{i,l}(k)D_{l,j}(k,\omega)\,
\label{resp}
\end{equation}
where $D^{(0)}_i(k,\omega)$ is the non--interacting particle--hole
propagator and ${\cal A}_{i,l}(k)$ is the Fourier transform of the
nucleon--nucleon effective interaction. The subscripts of the
various quantities take the values $1$ and $2$ for protons and
neutrons respectively. In this paper we use units such that
$\hbar=c=k_B=1$.
\par
In asymmetric nuclear matter isovector and isoscalar fluctuations
are coupled. However one can still separate oscillations with
neutrons and protons moving in phase (isoscalarlike) or out of phase
(isovectorlike) and add the contributions of the corresponding
variances. This can be done because the time scales of isoscalarlike
oscillations and of isovectorlike oscillations are very different
for nuclear matter with values of temperature and density
sufficiently close to the borders of the spinodal region. Indeed,
the growth rate of the unstable isoscalarlike modes, $\Gamma_k$,
turns out to be much smaller than the real frequencies,
$\omega^{iv}_k$, of isovectorlike modes \cite{Col05,Mat00}. Thus the
two kinds of fluctuations can be considered two independent
stochastic processes. In addition, for the relevant values of the
magnitude of the wave vector $k$, the following inequalities
$\Gamma_k/T< 1$ and $\omega^{iv}_k/T> 1$ hold. \par As a consequence
of the linear approximation for the stochastic field, the
probability distribution of density fluctuations,
$P[\delta\varrho_i({\bf k},t)]$, is given by a product of Gaussian
distributions:
\begin{equation}
P[\delta\varrho_i({\bf k},t)]=\,N\exp
\Big(-\frac{1}{2V}\sum_{\bf k}\sum_{i,j}
\delta\varrho_i({\bf k}^*,t)\big(\sigma^2(k,t)\big)^{-1}_{i,j}
\delta\varrho_j({\bf k},t)\Big)\, ,
\label{gauss}
\end{equation}
where $N$ is a normalization constant.
Each single factor corresponds to a stochastic process
for a given wave vector ${\bf k}$ \cite{Mat00,Mat03}.
The covariance matrix for the isoscalarlike fluctuations
\begin{equation}
\sigma^2_{i,j}(k,t)=TC^{is}_{i,j}(k)
\frac{1}{\Gamma_k}\Big(e^{2\Gamma_kt}-1\Big)\,,
\label{variance}
\end{equation}
can be evaluated taking the classical (thermodynamical) limit
$\omega/T\ll 1$ of the 
fluctuation--dissipation relation. The coefficients $C^{is}_{i,j}(k)$ are the
residues, times $i$, of the response function at the pole
$\omega=i\Gamma_k$ \cite{Col05}. \par
Since the time scale of the isovectorlike fluctuations is much shorter than
that of the growth of the unstable modes,
for the covariance matrix of the isovectorlike fluctuations we can take 
its asymptotic value for  $t\rightarrow\infty$, with nuclear matter at 
given values of density, temperature and asymmetry,
\begin{equation}
\sigma^2_{i,j}(k) = C_{i,j}^{iv}(k)\,,
\label{var_isov}
\end{equation}
where $C_{i,j}^{iv}(k)$ represent the residues of the response
function at the real pole $\omega^{iv}_k$. This equation has been
obtained by exploiting the fluctuation--dissipation theorem, now in
the limit $\omega/T>>1$ \cite{Col05}.  \par It should be noticed
that, in the low temperature limit, the isovector fluctuation  amplitude is
essentially different from the value expected at the thermodynamical
limit, $\omega<<T$, where the fluctuation variance is 
proportional to the temperature and inversely related to the value
of the symmetry energy \cite{Duc07}.
\par
With the aim to preserve a simple formalism, which, to some extent, allows us
to perform calculations analytically, we have evaluated the response function
of Eq. (\ref{resp}) within a semiclassical approximation (Vlasov equation).
However, this approximation cannot be extended to values of $k\geq k_F$
($k_F$ being the neutron or proton Fermi momentum).
We cure this shortcoming with an
{\it ad hoc} specific recipe. We calculate poles and residues of the
response function within the Vlasov approximation for $k\lesssim1.1\,k_M$,
where $k_M$ is defined by $\Gamma_{k>k_M}<0$. With the physical
parameters used in
this work $k_M$ is about the neutron Fermi momentum. For $k\geq1.1\,k_M$ we
take for the total variance of density fluctuation distributions 
its value for $k\rightarrow\infty$,
$\sigma^2_{i,j}(k,t)=\delta_{i,j}\varrho_i$ \cite{Walecka}, 
with $\varrho_i$ being the density of $i$--species nucleons. This recipe is
suggested by two circumstances. Firstly, for $k_M<k<1.1\, k_M$
the variance, which relaxes towards its asymptotic 
value since $\Gamma_k<0$, remains appreciably
larger than its limit $\varrho_i$ for $k\rightarrow\infty$.
Secondly, the latter is approached already for values of $k$ slightly 
larger than $1.1k_M$, as it can be shown by explicit calculations 
including quantum effects. The non--accurate evaluation of the 
response function in the interval $1.1\,k_M<k\lesssim 1.3\,k_M$ should 
not introduce sizeable effects, since the values of the variance in this
interval is about one order of magnitude smaller than that of the most
unstable mode.
\par
\subsection{Details of the interaction}
In the present paper we adopt the same schematic Skyrme--like effective
interaction as in Ref. \cite{Col05}:
\begin{equation}
{\cal A}_{i,j}(k)= {\cal A}(k)+{\cal S}_{i,j}(k)\,.
\label{interaction}
\end{equation}
For the symmetric term ${\cal A}(k)$ we use the finite--range effective
interaction introduced in Ref. \cite{ColA94}:
\begin{equation}
{\cal A}(k)=\Big(A\frac{1}{\varrho_{eq}}+(\sigma+1)\frac{B}
{\varrho_{eq}^{\sigma+1}}\varrho^{\sigma}\Big)e^{-c^2\,k^2/2}\,,
\label{inters}
\end{equation}
where $\varrho=\varrho_1+\varrho_2$ is the uniform mean value of
the total density of nucleons, $\varrho_{eq}=0.16~ {\rm fm}^{-3}$ is
the density of symmetric nuclear matter at saturation and
\[
A=-356.8\,{\rm MeV},~~B=303.9\,{\rm MeV},~~\sigma=\,\frac{1}{6}\,.
\]
The width of the Gaussian in Eq. (\ref{inters}) has been chosen in
order to reproduce the surface energy term as prescribed in Ref. \cite{Mye66}.
\par
The isospin--dependent part, ${\cal S}_{i,j}(k)$, contains three different
terms
\begin{equation}
{\cal S}_{i,j}(k)=\frac{\partial^2{\cal E}_{sym}}{\partial\varrho_i
\partial\varrho_j}+\tau_i\tau_jDk^2+\frac{1+\tau_i}{2}V_C(k)\delta_{i,j}
\,,
\label{interv}
\end{equation}
with $\tau_1=1$ and $\tau_2=-1$. Here ${\cal E}_{sym}$ represents
the potential part of the symmetry--energy density. 
For the coefficient of the isovector surface term we use the value
$D=40\,{\rm MeV\,fm}^5$ \cite{Bay71}. Moreover, we include the
Coulomb interaction $V_C(k)$ according to the approach of Ref. \cite{Fab98}. 
A mean--field exchange contribution
\[V_C^{ex}=-\frac{1}{3}\Big(\frac{3}{\pi}\Big)^{1/3}e^2\varrho_1^{-2/3}
\]
is also added to the bare Coulomb force.
\par
In order to stress the effects of the asymmetry of the nuclear
medium, we will present results obtained with two different
parametrizations of the symmetry energy: one with a stronger density
dependence (``superstiff'' asymmetry term) and the other one with
a weaker density dependence (``soft'' asymmetry term). In both
cases the density dependence of the potential part of the
symmetry--energy density can be expressed by
\begin{equation}
{\cal E}_{sym}(\varrho_1,\varrho_2)=S(\varrho)(\varrho_2-\varrho_1)^2
\,,
\label{symmetry}
\end{equation}
with
\begin{equation}
S(\varrho)=\frac{2d}{\varrho_{eq}^2}\frac{\varrho}{1+\varrho/\varrho_{eq}}
\,,
\label{stiff}
\end{equation}
where $d=19\,{\rm MeV}$ \cite{Bao00}, for the ``superstiff'' case, and
\begin{equation}
S(\varrho)=d_1-d_2\varrho
\,,
\label{soft}
\end{equation}
where $d_1=240.9\,{\rm MeV\,fm}^3$ and $d_2=819.1\,{\rm MeV\,fm}^6$
\cite{ColA98}, for the ``soft'' case. It should be noticed that
$S(\varrho)$ is nothing but the potential part of the 
symmetry--energy coefficient divided by $\varrho$, $S(\varrho) =
C^{pot}_{sym}(\varrho)/\varrho$. The inclusion of the Coulomb interaction
gives rise to an overall decrease of the growth rate of density
fluctuations with a corresponding contraction of the instability
region in the ($\varrho,T$) phase diagram \cite{Fab98,Col02}.
\par
\subsection{Isospin effects}
Proton and neutron densities oscillate in phase and out of phase,
respectively in the isoscalarlike fluctuations and in the
isovectorlike fluctuations, although with different amplitudes in
general. The ratio between amplitudes is given by
$\sigma^2_{1,1}(k)/ \sigma^2_{1,2}(k)=\pm\sqrt{(\sigma^2_{1,1}(k)/
\sigma^2_{2,2}(k))}$, with $+$ for the isoscalarlike case and with
$-$ for isovectorlike case. The above relation follows from the 
relevant property 
\begin{equation}
{\rm det}|C_{i,j}^{is,iv}(k)|=0\,,
\label{det}
\end{equation}
of the residues. Hence, for asymmetric matter, even in unstable 
isoscalarlike oscillations that lead to phase separation, protons 
and neutrons move with different amplitude. In particular one 
observes that the ratio between proton and neutron fluctuations is
larger than the $Z/N$ ratio of the original matter, leading to a
more symmetric liquid phase, the so--called isospin distillation
effect \cite{Mue95,Bao97,Bar98,Rep04,Rep05,Xu00}. The ratio between 
the proton and neutron density 
fluctuations is mostly determined by the isoscalarlike fluctuations.
This is simply related to the fact that during the spinodal
decomposition, the isoscalarlike fluctuations are much larger than
the isovectorlike ones. Thus, this ratio, for a given value of $k$,
can be put as:
\begin{equation}
\frac{\delta\varrho_1({\bf k})}{\delta\varrho_2({\bf k})}
\simeq\,\sqrt{\frac{C^{is}_{1,1}(k)}{C^{is}_{2,2}(k)}}
\,,
\label{}
\end{equation}
where the ratio of the residues at the imaginary pole, $i\Gamma_k$, of
$D_{i,j}(k,\omega)$ is given by
\begin{equation}
\frac{C^{is}_{1,1}(k)}{C^{is}_{2,2}(k)}=
\frac{D^{(0)}_1(k,i\Gamma_k)\big(1-D^{(0)}_2(k,i\Gamma_k){\cal A}_{2,2}(k)
\big)}
{D^{(0)}_2(k,i\Gamma_k)\big(1-D^{(0)}_1(k,\omega){\cal A}_{1,1}(k)\big)}
\,.
\label{ratiore1}
\end{equation}
The non--interacting particle--hole propagator, in the semiclassical
approximation, is expressed by \cite{Col05}
\[
D_{i}^{(0)}(k,i\Gamma_k)\simeq -\frac{\partial \varrho_i}{\partial \tilde\mu_i}
+\frac{1}{2\pi}m^2F(\beta \tilde\mu_i)\frac{\Gamma_k}{k}\, ,
\]
where the effective chemical potential $\tilde\mu_i$ of neutrons or
protons is measured with respect to the uniform mean field
$U_i(\varrho_1,\varrho_2)$ of the unperturbed initial state, and
$F(\beta \tilde\mu_i)$ is the function 
\[F(\beta \tilde\mu_i)=\,\frac{1}{e^{-\beta \tilde\mu_i}+1}\,,\]
with $\beta =1/T$ being the inverse temperature. For the physical
parameters considered in the present paper the second term in the expression
of $D_{i}^{(0)}(k,i\Gamma_k)$ can be neglected at a satisfying approximation,
then Eq. (\ref{ratiore1}) becomes
\begin{eqnarray}
\frac{C^{is}_{1,1}(k)}{C^{is}_{2,2}(k)}=\frac{{\displaystyle
\frac{\partial \tilde\mu_2}
{\partial \varrho_2}}+{\cal A}_{2,2}(k)}{{\displaystyle
\frac{\partial \tilde\mu_1}
{\partial \varrho_1}}+{\cal A}_{1,1}(k)}
=\frac{{\displaystyle\frac{\partial^2f}{\partial\varrho_2^2}}
+{\cal A}_{2,2}(k)}{{\displaystyle\frac{\partial^2f}{\partial \varrho_1^2}}
+{\cal A}_{1,1}(k)}\, ,
\label{ratiore2}
\end{eqnarray}
where $f=f(\varrho,\varrho_3)$ with $\varrho_3=\varrho_2-\varrho_1$,
is the sum of the kinetic and entropy terms of the free--energy
density, i.e. $f$ represents the free--energy density of a
non-interacting two--component Fermi gas with effective chemical
potentials $\tilde\mu_i$. \par For a qualitative analysis we can
neglect the Coulomb interaction. For small asymmetry values,
we can limit ourselves to consider only
the first order term in the expansion of Eq. (\ref{ratiore2}) in
powers of the asymmetry $\alpha=\varrho_3/\varrho$. In this case the
ratio of residues is given by
\begin{equation}
\frac{C^{is}_{1,1}(k)}{C^{is}_{2,2}(k)}\simeq 1+\frac{1}{{\cal B}(k)}
\Big(4\frac{\partial}{\partial\varrho}\frac{\partial^2f(\varrho,\varrho_3)}
{\partial\varrho_3^2}|_{\varrho_3=0}+
8\frac{\partial S(\varrho)}{\partial\varrho}\Big)\varrho\alpha\, ,
\label{distil}
\end{equation}
where ${\cal B}(k)$ represents the sum of the part of interaction of
Eq. (\ref{interaction}) common to both the species of nucleons, and of
the second derivatives of the free--energy density
$f(\varrho,\varrho_3)$:
\[{\cal B}(k)={\cal A}(k)+2S(\varrho)+D\,k^2+\frac{\partial^2
f(\varrho,\varrho_3)}{\partial\varrho_3^2}|_{\varrho_3=0}+
\frac{\partial^2 f(\varrho,0)}{\partial\varrho^2}\, .\] Equation
(\ref{distil}) shows that for the isospin distillation effect a
crucial role is played by the derivative of the coefficient of the
symmetry energy with respect to the total density. \par With the
parameters of the interactions considered here, we get, for systems
moderately inside the spinodal region, a larger distillation effect
with the ``superstiff'' interaction. However, it should be remarked
that the derivative of $S(\varrho)$ depends on the density. Actually
at rather low  densities  distillation effects are expected to be
stronger with the "soft" parametrization \cite{Bar02}.
\par
The symmetry energy also plays an important role in determining
isovector fluctuations, $\delta\varrho_3$. These lead to
fluctuations in the $(N-Z)$ content of a given density domain of mass
$A = N+Z$. For finite values of the asymmetry $\alpha$, it is
generally not easy to single out from the isospin--dependent
interaction, $S_{i,j}(k)$, a term which can play a conclusive role
in determining the widths of isotopic distributions. Only for
$\alpha=0$ and, in addition, neglecting the Coulomb interaction,
isovector and isoscalar fluctuations are decoupled. In this case the
fluctuations $\delta\varrho_3$ are only due to the isovector modes,
and the residues at the real pole $\omega^{iv}_k$ are solely
determined by the coefficient $S(\varrho)$ in the expression of the
symmetry energy of Eq. (\ref{symmetry}). For finite values of
$\alpha$, terms containing derivatives of $S(\varrho)$ contribute to
the magnitude of the residues $C_{i,j}^{iv}(k)$ as well. 
Generally speaking, one can expect that the strength of the
isovector pole increases with the coefficient $S(\varrho)$.
Hence, in the limit considered here, $\omega / T >> 1$, isovector
fluctuations becomes larger when increasing the symmetry 
energy coefficient.
It should be noticed that also for $\alpha\not =0$ the contributions
to $\delta\varrho_3$ of the isoscalarlike modes still tend to cancel out. 

\subsection{{\bf Cluster} recognition}

We consider a finite cubic lattice of volume $V$. We assume the
density vanishing at the surface of the box. The number of nucleons
is then fixed and in the box the liquid and vapour phases coexist.
Moreover, finite size effects can be  taken into account in this
way. However, it should be remarked that the condition of vanishing
density at the border of the box imposes some symmetries to the
problem. In fact, in this case the number of independent
oscillations is reduced by $2^3$. Accordingly, due to the imposed
symmetries, also fluctuations variances are reduced. This is not so
important in the isoscalarlike case, since these modes are unstable
and fluctuations are amplified anyway. On the contrary,
isovectorlike fluctuations remain quenched (by a factor $8$).
However, this does not affect our conclusions about the relative comparison
of the results obtained with the different parametrizations of the
symmetry energy. 
\par
In order to extract the spatial density $\delta\rho_i({\bf r})$, we
perform a Fourier expansion of the fluctuations $ \delta\rho_i({\bf
k})$. The Fourier components contain the product of three sine
functions:
\begin{equation}
\phi_{\bf k}({\bf r})=\Big(\frac{2}{L}\Big)^3\,
\sin(k_1x)\sin(k_2y)\sin(k_3z)\,, \label{}
\end{equation}
with $k_i=2\pi\,n_i/L$ and $x=n_xb$, $y=n_yb$, $z=n_zb$ ($n_i$ and
$n_{x,y,z}$ take positive integer values), where $L$ and $b$ are the
lengths of the sides of the box and of the primitive cell respectively. 
The coefficients of the functions $\phi_{\bf k}({\bf r})$ 
in the Fourier expansion  are linear combinations 
of the isoscalarlike and isovectorlike 
Fourier coefficients $\delta\varrho_i({\bf k},t)$, given by stochastic
processes with the probability distribution of Eq. (\ref{gauss})
and the ratio of proton to neutron amplitudes given by
$\sigma^2_{1,1}(k)/ \sigma^2_{1,2}(k)$.
\par
The size of the box will be chosen in order that the box contains a
number of nucleons of the same order of magnitude as in actual heavy
ion collision experiments. While the adopted size of the primitive
cell, $V_{cell}$,  is such that at least a nucleon of 
both the species is present 
in the cell on the average. Clusters of nucleons are associated 
with large density domains and will be formed by means of a coalescence
algorithm, solely based on the density of neighbouring
cells. Adjacent cells with a value of the density above (liquid)
or below (vapour) the average density are collected together. 
Then, we obtain domains of higher density surrounded by domains of 
lower density. In this way we can investigate separately the 
properties of the liquid phase and of the vapour phase. Moreover, 
we can build observables
starting from event by event cluster distributions. 
\par

\section{\label{BB}Results}
\begin{figure}
\includegraphics{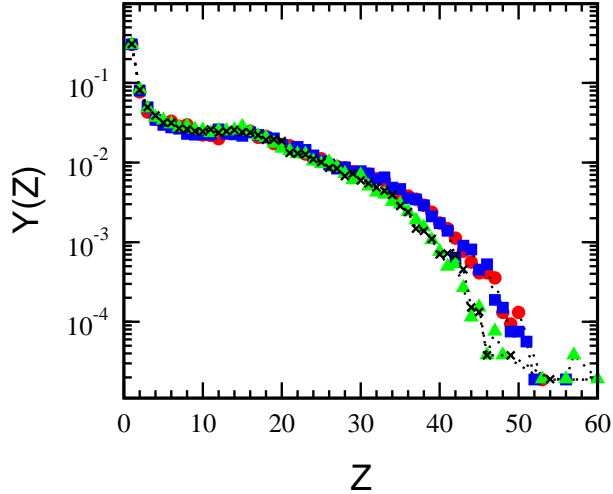}
\caption{\label{fig1}(Color online) Charge distributions 
for clusters belonging to 
the high density phase for two values of the global asymmetry
$\alpha_0$, calculated with the ``soft'' asymmetry term, circles 
($\alpha_0=0.1$) and triangles ($\alpha_0=0.2$), and the
``superstiff'' asymmetry term, squares ($\alpha_0=0.1$) and
crosses ($\alpha_0=0.2$). }
\end{figure}
\begin{figure}
\includegraphics{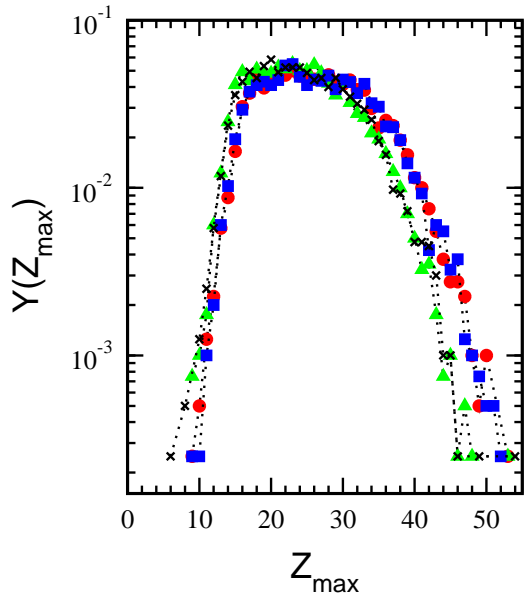}
\caption{\label{fig2} (Color online) Distributions of the heaviest cluster
for two values of the global asymmetry $\alpha_0$. Symbols as in Fig.
\ref{fig1}.
}
\end{figure}

In this section we present and discuss statistical distributions of
the clusters obtained using the coalescence recipe outlined above.
In particular we focus on the isospin content of the clusters coming
from both the liquid phase and the vapour phase. The value chosen
for the average density $\varrho=0.4\varrho_{eq}$ and the
temperature $T=4.5\,{\rm MeV}$ are in the range expected for the
multifragmentation process \cite{Rep04,Fra01}. For the time that the
system spends in the instability region, we have chosen a value of
$t=80 {\rm fm/c}$. This value is compatible with that obtained
within the stochastic mean--field approach of Ref. \cite{Bar02}.
Moreover, in this short time interval the growth of fluctuations is
still limited so that a linear approximation can be considered as
reasonable. For the side of the cubic lattice containing the nuclear
system, we have adopted the value $L=21\,{\rm fm}$, while the side
of the primitive cell has a length of $3\,{\rm fm}$. With the chosen
value of density the box contains $\simeq 370$ nucleons. 
\par 
We have performed calculations for the two parametrizations of the symmetry
energy introduced above and for two values of the global asymmetry,
$\alpha_0 = 0.1$, $0.2$. We have run $4\cdot 10^3$ events for each
case. With the chosen values of the parameters the average numbers
of nucleons in the liquid phase and in the vapour phase are $\simeq
270$ and $\simeq 100$ respectively. \par
\begin{figure}
\includegraphics{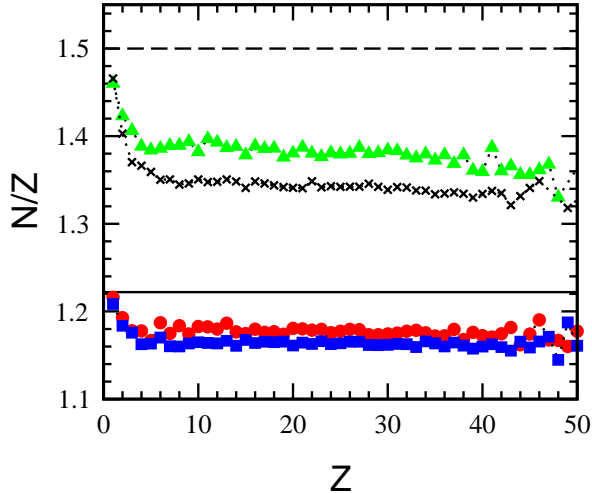}
\caption{\label{fig3} (Color online) Average ratio $N/Z$ as a 
function of $Z$ for 
clusters belonging to the high density (liquid) phase, calculated
with the ``soft'' asymmetry term (circles and triangles) and the
``superstiff'' asymmetry term (squares and crosses). Bottom:
global asymmetry $\alpha_0=0.1$. Top: $\alpha_0=0.2$. Solid
horizontal line: global ratio $N/Z$ for $\alpha_0=0.1$. Dashed
horizontal line: global ratio $N/Z$ for $\alpha_0=0.2$. }
\end{figure}
\begin{figure}
\includegraphics{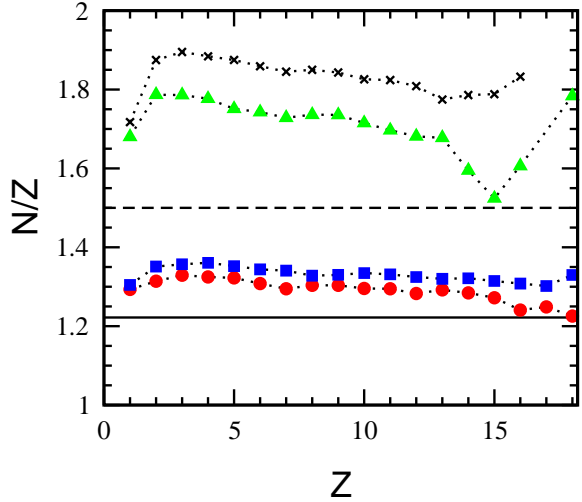}
\caption{\label{fig4} (Color online) The same as in Fig.~\ref{fig3} 
but for clusters belonging to the low density (vapour) phase.
}
\end{figure}

\subsection{Charge distributions}
In order to assess the validity of our approach, we have evaluated the 
distribution of the cluster yield. In Fig.~\ref{fig1} we 
present the charge distribution of clusters belonging to the liquid
phase, calculated for two values of the global asymmetry ($\alpha_0
= 0.1,0.2$) and for the two parametrizations of the symmetry energy
considered. One can see that the four curves displayed in the Figure
are rather similar. This is expected from the fact that
isoscalarlike modes and related variances, that determine the
density growth and the appearance of fragments, do not much depend 
on the behaviour of the symmetry energy, nor on the global asymmetry
of the matter considered. Moreover, the obtained charge distribution
is similar to the results of full stochastic mean--field simulations
\cite{Rep04,Fra01} of heavy ion collisions at $\simeq 30 MeV/A$, where
similar conditions of temperature and density inside the spinodal
region are encountered. This indicates that fragment size is 
essentially determined by the properties of the most unstable normal
modes, as derived in the linear approximation. The beating of these
modes leads to the rather wide charge distribution. The charge
distribution of Fig.~\ref{fig1} compares rather well also with the
experimental results of Ref. \cite{Fra01}, where fragmentation of
systems of similar size is investigated.

Another observable of experimental interest is 
the distribution of the heaviest cluster obtained in each event.
The study of this observable requires, of course, an event by event analysis.
Results are shown in Fig.~\ref{fig2}. Also in this case we observe
that the four curves corresponding to different charge asymmetries and to
different parametrizations of the symmetry energy look quite similar.
Moreover, our results are similar to the predictions of full
simulations and to experimental data \cite{Fra01}.  

\subsection{Isotopic properties}

From Figs.~\ref{fig3} and \ref{fig4} we can appreciate the isospin
distillation process occurring in the spinodal decomposition of 
asymmetric nuclear systems. The average ratio $N/Z$ is plotted as a
function of $Z$ for the correlated density domains of the liquid
phase (see Fig.~\ref{fig3}) and of the vapour phase (see 
Fig.~\ref{fig4}). One can nicely see that the ratio $N/Z$ is
smaller than the initial global value for the liquid, while the
opposite holds for the vapour phase. We notice that this effect is 
present also for the two coexisting phases at thermodynamical 
equilibrium, as shown in Refs. \cite{Rep05,Duc07}. Moreover, 
we observe a  slight decrease of the ratio 
$N/Z$ with $Z$. This trend is essentially related to the 
fact that lighter clusters originate 
from lower density domains where, as discussed in the following (see 
Subsection C), isospin distillation is less effective.  We notice that
this behaviour, which is in agreement 
with the results of previous calculations based on dynamical models 
\cite{Rep05,Liu04}, is not observed in statistical model calculations, 
that exhibit an opposite trend \cite{Bot01,Lar99}.
Hence, this property of the ratio $N/Z$ with respect to the cluster charge
would allow to disantangle between the predictions of different models. \par
In particular, Figs.~\ref{fig3} and 
\ref{fig4} show that isospin distillation increases with 
increasing the global asymmetry and is larger when 
the ``superstiff'' asymmetry term is used with 
respect to the ``soft'' case. These features are explained by 
Eq. (\ref{distil}). Indeed, the induced asymmetry is proportional to the
global asymmetry $\alpha_0$ and, at the considered density, the
derivative $\partial S(\varrho)/\partial \varrho$ has nearly the
same magnitude, but opposite sign for the two used parametrizations
of the symmetry energy (positive for the ``superstiff'' case and
negative for the ``soft'' case). 
\par
\begin{figure}
\includegraphics{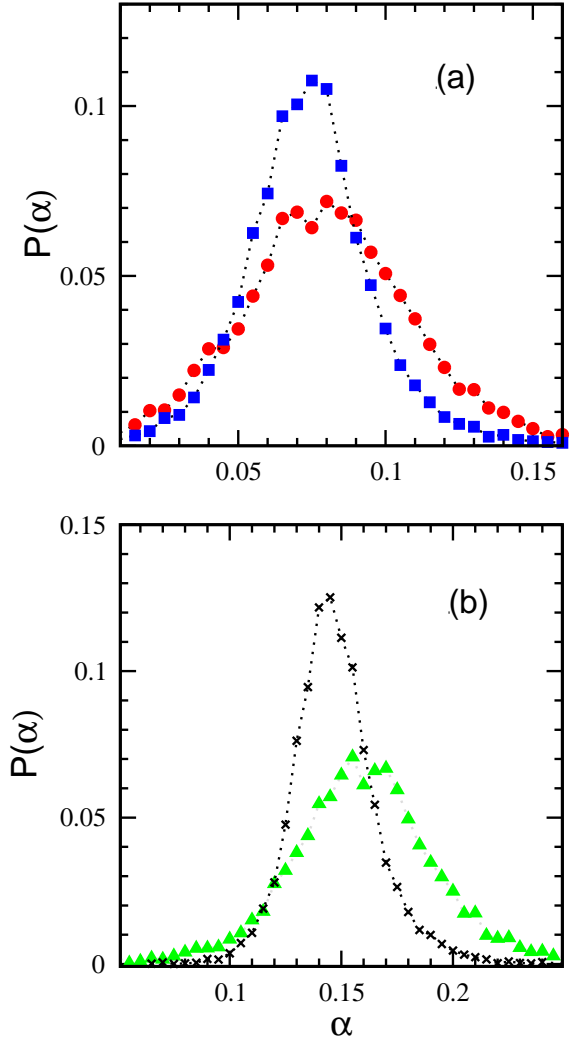}
\caption{\label{fig5} (Color online) Average distribution of 
asymmetry $\alpha$  for
clusters belonging to the high density (liquid) phase, calculated
with the ``soft'' asymmetry term (circles and triangles) and the
``superstiff'' asymmetry term (squares and crosses). Panel (a): 
global asymmetry $\alpha_0=0.1$. Panel (b): $\alpha_0=0.2$. }
\end{figure}
\begin{figure}
\includegraphics{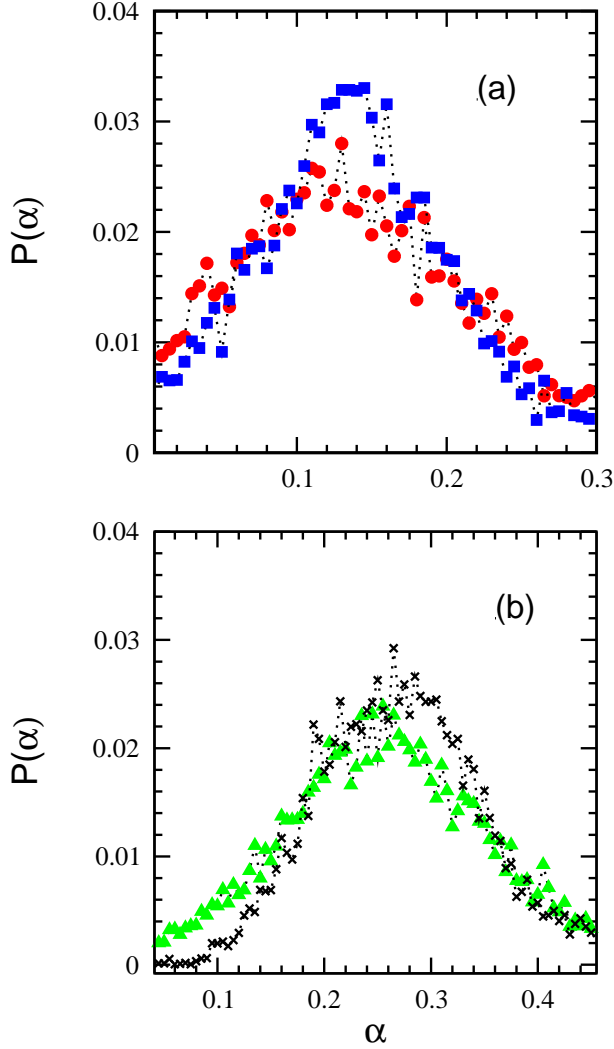}
\caption{\label{fig6} (Color online) The same as in Fig.~\ref{fig5} 
but for clusters belonging to the low density (vapour) phase.
}
\end{figure}

Now we turn our attention to the probability distribution of the
asymmetry parameter for the domains belonging to the two phases. In
Figs.~\ref{fig5} and \ref{fig6} we report the probability of finding
a cluster with the asymmetry $\alpha=(N-Z)/(N+Z)$ for the liquid and
vapour phases respectively. The distribution is averaged over
intermediate mass clusters ($5<N+Z<50$) for the high density phase
and over light clusters ($3<N+Z<15$) for the low density phase.
The shift of the maximum toward the left (right) for the liquid 
(vapour) phase in the ``superstiff'' case with respect to the
``soft'' case is due to the more effective distillation power of the
``superstiff'' interaction, as already seen above. In addition we
observe that distributions are wider in the ``soft'' case. This 
effect can be related to the fact that at the considered density the
coefficient $S(\varrho)$ of Eq. (\ref{symmetry}) is larger
in the "soft" case with respect to the "superstiff" case, leading 
to a larger value of the residues in Eq. (\ref{var_isov}).
Moreover, the distributions are generally broader for the vapour
phase. In the following, we will show that this feature can be
easily explained on the basis of the relative contributions of
isoscalarlike and isovectorlike oscillations, at least within our
linear treatment of fluctuations. Moreover, we observe that the
asymmetry distributions become narrower with increasing the global
asymmetry for both the two phases. This was observed also 
in Ref. \cite{Duc07}, within a thermodynamical study of spinodal
decomposition. A final remark on this topic is in order. 
As already seen for the distillation effect, the behavior of the
width of the isotopic distributions, in connection with the parametrization
adopted for the symmetry energy, depends on the density value of the
matter undergoing spinodal decomposition. With the
present approach the widths, though slightly, result to be smaller
in the ``superstiff'' case than in the ``soft'' case, whereas in
Ref. \cite{Col05} we have observed an opposite tendency. 
In fact, with the choice of parameters of Ref. \cite{Col05} nuclear
matter explored a region of higher instability (lower density) 
with respect to the present calculations. Then, the ratio between
the amplitudes of the isoscalarlike fluctuations (that also
contribute to the isotopic variance, due to the dependence of the
distillation effect, $\delta\varrho_2/\delta\varrho_1$, on the wave
number $k$) and of the isovectorlike fluctuations was quite larger
than here. Furthermore, the weighting procedure of the different
fluctuation modes in Ref. \cite{Col05} could likely underestimate
the isovectorlike fluctuations. \par

\subsection{Density dependence of distillation effect}

Now we show  that the amplitude of the isospin distillation effect, as
well as the variances of the isotopic distributions, is related to
the value of the density reached in the different domains associated
with nucleon clusters.\par
Due to the beating of the several unstable
modes, clusters with a given charge are not associated with a fixed
density domain, but may come from domains of different volume having
different density, i.e. the cluster density $\varrho$ may
fluctuate. We will show that this effect influences the cluster 
isotopic properties.

\begin{figure}
\includegraphics{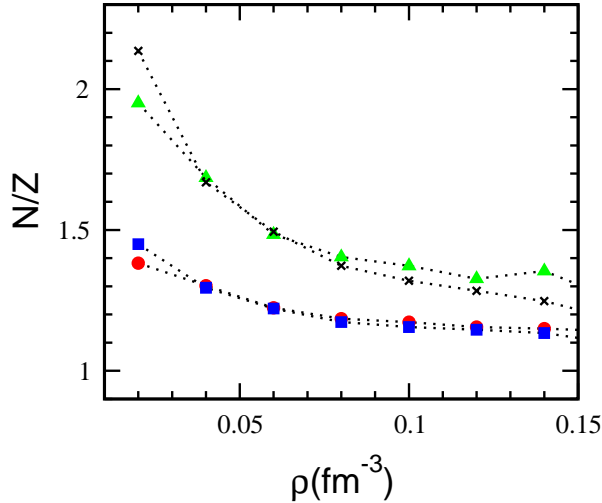}
\caption{\label{fig7} (Color online) Average ratio $N/Z$ as a function of the
cluster density for two values of the global asymmetry $\alpha_0$,
calculated with the ``soft'' asymmetry term, circles 
($\alpha_0=0.1$) and triangles ($\alpha_0=0.2$), and the
``superstiff'' asymmetry term, squares ($\alpha_0=0.1$) and
crosses ($\alpha_0=0.2$) }
\end{figure}

As far as the distillation effect is concerned, we notice that the $N/Z$
ratio of a cluster can be written as:
$$ N/Z = (\varrho_2^0 + \delta\varrho_2) / (\varrho_1^0 + \delta\varrho_1), $$
where $\varrho_2^0$ and $\varrho_1^0$ are the neutron and proton
densities of the initial matter, while $\delta\varrho_2$ and
$\delta\varrho_1$ are the corresponding fluctuation values in the
domain considered. After some algebra one obtains:
$$N/Z = (N/Z)_0 -  \frac {[(N/Z)_0 -
\delta\varrho_2 / \delta\varrho_1] \cdot \delta\varrho_1 / \varrho_1^0 }
{1 + \delta\varrho_1 / \varrho_1^0}
$$
where $(N/Z)_0 = \varrho_2^0 /
\varrho_1^0$. We notice that, due to the distillation effect,
$\delta\varrho_2 / \delta\varrho_1 < (N/Z)_0$. From the expression
given above, one can see that the ratio $N/Z$ of a given cluster 
decreases when the density $\varrho$ of the domain increases,
i.e. $\delta\varrho_1$ (and $\delta\varrho_2$) are larger. In
fact, larger densities are associated with larger isoscalarlike
fluctuations and thus with a larger distillation. On the other hand,
clusters in the vapour phase, having density smaller than the
initial one ($\delta\varrho < 0$), show a larger value of the
ratio $N/Z$. Moreover, the deviation of the ratio $N/Z$ with respect
to the initial  value $(N/Z)_0$ increases as far as the ratio
$\delta\varrho_2 / \delta\varrho_1$ gets smaller, i.e. for
parametrizations of the symmetry energy that lead to larger
distillation effect. In Fig.~\ref{fig7} we show the $N/Z$ ratio, 
averaged over all charges, as a function of the cluster density. 
The decreasing trend is clear, especially for the more neutron--rich
system, where one can also better appreciate the different effect of
the two equations of state employed in the calculations. 
\par
For the variance of the isotopic distributions, we also expect a
decreasing behaviour with density. In fact isovectorlike
fluctuations, that are mainly responsible of the isotopic variance,
in our linear approach are implemented according to 
the properties (density, asymmetry, temperature) of the
initial matter. Then, nucleon clusters with larger density should have
narrower isotopic distributions due to the larger contribution of
the isoscalarlike modes with respect to isovectorlike fluctuations. 
\par
\begin{figure}
\includegraphics{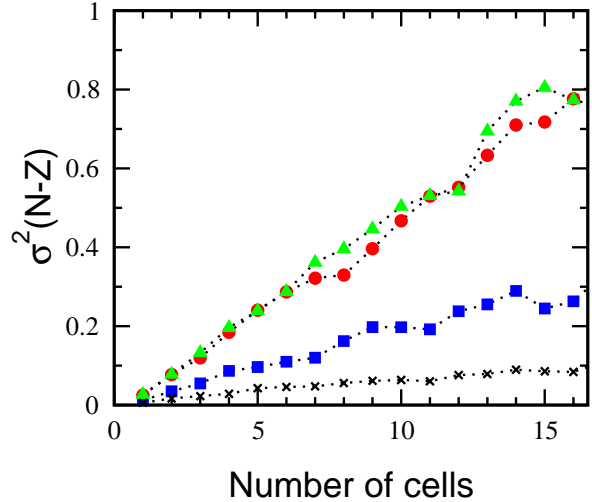}
\caption{\label {fig8} (Color online) Average variance of the ($N-Z$) 
distribution for clusters of volume $V=N_VV_{cell}$ as a function of 
the number of the elementary cells $N_V$ calculated for two values of the
global asymmetry $\alpha_0$. Symbols as in Fig. \ref{fig7}.}
\end{figure}

The probability distribution for the isovector density fluctuations
$\delta\rho_3$ in a given domain of volume $V$ can be
approximatively expressed as:
\begin{equation}
P \approx \exp(-\delta\rho_{3}^2/2\sigma_{iv}^2),
\label{white}
\end{equation}
where the variance $\sigma_{iv}^2$  can be evaluated starting from
the value of $\sigma^2_{ij}(k)$ of Eq. (\ref{var_isov}) and is inversely
proportional to the volume $V$, $\sigma_{iv}^2 = g(\varrho,T)/V$,
where $g(\varrho,T)$ is a function that depends on the initial
conditions of the matter undergoing spinodal decomposition and on
the parameters on the interaction adopted \cite{EPJA06}. Then, for 
a cluster of volume $V$, the distribution $P(N-Z)$ can be written as:
\begin{equation}
P(N-Z) \approx \exp\Big(-[N-Z-({N_0}-{Z_0})]^2/
[2V g(\varrho,T)]\Big).
\label{gauss2}
\end{equation}
Hence, the variance of the isotopic distribution should
be proportional to $V$.  This is confirmed by our calculations, as
shown in Fig.~\ref{fig8}, where we plot the variance of the $(N-Z)$
distribution, averaged over all clusters with  volume $V = N_V V_{cell}$, as
a function of $N_V$. From Fig.~\ref{fig8} one can also see that, 
especially in the case of the ``superstiff'' interaction, the 
variance is smaller  for the neutron richer system (compare squares 
and crosses). Moreover, for a given volume, variances are larger in 
the "soft" case. 
\par
If we consider now clusters with a given mass $A$, on can easily realize 
that the isotopic distributions will depend on the cluster density,
being broader for clusters with smaller density (i.e. larger volume).
Hence clusters that originate from larger density domains, 
are expected to have an average  $N/Z$ ratio
smaller than the value corresponding to the initial matter and,
at the same time, are expected to show narrower isotopic distributions.

\section{Conclusions}

In this paper we have considered a piece of asymmetric nuclear
matter contained in a cubic box with impenetrable walls. The
values of density and temperature of the system are inside the
spinodal instability region of the phase diagram. We have mainly
focused our attention on the isotopic distributions of clusters
coming from the break--up of the system.
\par
In our approach the density fluctuations are considered within a
linear approximation, then our results refer to nuclear matter
being not deeply inside the unstable spinodal region. Such physical
situation can occur in central collisions of heavy ions at moderate 
energy ($\simeq 30 MeV/A$). Moreover, we remark that our 
investigation concerns the distributions just after the early
break--up of the nuclear system. 
\par 
It has been found that, at the considered values of density 
and asymmmetry, the global features of 
the disassembly, the isospin distillation effect included, are
essentially determined by the unstable isoscalarlike fluctuations of
the density, while the widths of the isotopic distributions are
mostly  affected by the isovectorlike oscillations.
\par 
Within the model developed it is possible to recognize clusters 
originating from domains having different density values.
Discerning clusters with density lower or
higher than the initial density, we have observed a clear occurrence
of the isospin distillation effect: low--density clusters are more
neutron--rich than the initial system, correspondingly the opposite
happens for higher density clusters. As a general trend the average
value of the $N/Z$ ratio decreases with increasing the cluster
density. This effect results to be enhanced for more
charge--asymmetric systems. Moreover, it has been found that the
isotopic distributions of clusters belonging to the low density
phase are broader than those of clusters belonging to the high
density phase. Hence larger distillation effects are associated 
with narrower isotopic distributions, at least within our linear 
treatment of fluctuations. This feature, if it survives the 
secondary decay of primary fragments, could be qualitatively checked 
in experiments by  comparing the isotopic distributions of 
light clusters, that should emerge from the low-density phase,
and intermediate mass fragments, most likely associated with higher
density regions. However, at the present level of investigation,  
a careful comparison with experiments would require a reconstruction 
of the properties of the primary fragments, based on actual 
experimental data \cite{Hud03}. 
\par 
In order to stress the sensitivity of the results to the 
isovector part of the nuclear interaction,
we have presented calculations obtained with two different
parametrizations of the symmetry energy. For the values of
charge--asymmetry considered, we have seen that isospin effects are
essentially related to the values of the coefficient of the
symmetry energy and of its derivative with respect to the density.
In particular, the former quantity mainly determines the width
of the isotopic distributions, while the latter quantity plays a
crucial role for determining the strength of the isospin
distillation. However, it should be remarked that the symmetry energy 
properties depend on the density reached by the system
during the decomposition process. Then, the effectiveness of a given
parametrization of the symmetry energy for the isospin distillation
depends on the actual path followed by the nuclear system in the
density--temperature phase diagram.

\end{document}